# On Photoionization in the Hard X-Ray Region

**M. Ya. Amusia[1, 2][1], L. V. Chernysheva[2], and V. G. Yarzhemsky[3]**

[1] Racah Institute of Physics, the Hebrew University, Jerusalem, 91904, Israel
[2] Ioffe Physical-Technical Institute, RAS, St. Petersburg, 194021, Russia
[3] Kurnakov Institute of General and Inorganic Chemistry, RAS, Moscow, 119991, Russia

**Abstract**

It is demonstrated that recent experiments on ionization of 3p and 3s electrons in Ni film and solid body [1] cannot be described in the frames of Hartree-Fock or the random phase approximation with exchange. The deviation is so big that from the theoretical side it requires inclusion of huge and yet not entirely understood effects. Perhaps, it requires also experimental efforts in order to clarify the prominent difference obtained for films and crystal samples.

**1. Introductory remarks.** Recently the first measurements were performed of the ionization cross-section by photons in the energy range $\omega = 2 - 9keV$ on 3s and 3p subshells of Ni thin films and crystals [1]. The results were compared to old calculations [2], and an essential difference, particularly for the 3s subshell, in the high frequency region was disclosed. It appeared that the measured cross-section is smaller than the calculated one by a factor 2.5-3.

An essential difference was observed also between thin films and crystals, which is surprising since the ionization potentials of 3s and 3p subshells are much bigger than the typical solid state binding energies and much smaller than $\omega$.

These experiments are of special interest since until recently our knowledge on the high $\omega$ photoionization cross-section was grounded only upon theoretical considerations. Moreover, it was a general belief that with $\omega$ growth the cross-section becomes hydrogenlike, with its textbook asymptotic behavior $\sigma_{nl}(\omega) = A_{nl} / \omega^{l+7/2}$, where $nl$ are the principal quantum number and angular momentum of the ionized subshell.

**2. Main formulas.** It was demonstrated in [3] that the corrections of the random phase approximation with exchange (RPAE) remain important even in the high photon energy limit, so the simple presented above hydrogenlike formula is non-valid. It was shown there that the RPAE corrections to all but s-subshells are non-negligible at any high, but non-relativistic values of $\omega$. So, it was quite natural to apply the high $\omega$ RPAE approach developed in [3] in an attempt to reproduce the data from [1]. As was shown in [3] the RPAE at high $\omega$ is considerably simplified reducing to the following diagrammatic equation, where the shaded circle denotes the photoionization amplitude. The dashed line presents a photon, while a line with an arrow to the right (left) stands for an excited electron (vacancy) and a wavy line presents the interelectron Coulomb interaction (e.g. [4]):

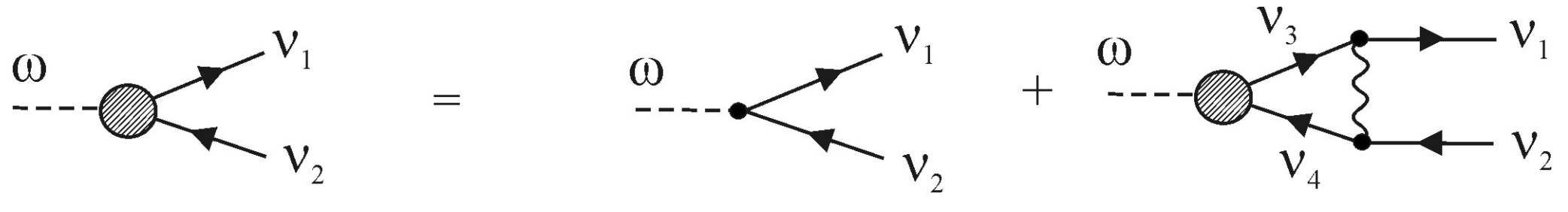

a) b) c) (1).

[1] E-mail: amusia@vms.huji.ac.il

Analytically, the corresponding to (1) equation looks like

$$\left\langle \nu_1 \middle| D(\omega) \middle| \nu_2 \right\rangle = \left\langle \nu_1 \middle| \hat{d} \middle| \nu_2 \right\rangle - \sum_{\nu_3 \geq F, \nu_4 < F} \frac{\left\langle \nu_3 \middle| D(\omega) \middle| \nu_4 \right\rangle \left\langle \nu_4 \nu_1 \middle| V \middle| \nu_3 \nu_2 \right\rangle}{\varepsilon_{\nu_4} - \varepsilon_{\nu_3} + \omega + i\eta}, \tag{2}$$

where $D(\omega)$ and $\hat{d}$ are RPAE and one-electron photoionization amplitudes, respectively, $\nu \geq F$ denotes summation and integration over excited and $\nu < F$ - summation over occupied HF one-electron states, $V$ is the Coulomb interelectron interaction, $\eta \to +0$.

At high $\omega$ only the exchange term, namely (1c), is taken into account in the right hand side, while the so-called direct and time-reverse terms have to be neglected [3]. The equation (2) has been solved numerically, using modified version of codes described in [5].

Exact solution of RPAE equations for unfilled shell atoms like Ni requires consideration of interaction of a large number of terms appearing after photoionization. In this Letter the exact solutions of RPAE equations was replaced by configuration average approximation. This approximation makes possible to estimate general trends of electron correlation effects independent of a particular choice of the ground term. In this approximation the weight factors for the interaction between two open shells are obtained by multiplying the weight factors for filled shells by normalization factor $(N_1 N_2)/[(4l_1{+}2)(4l_2{+}2)]$, where $N_{1,2}(l_{1,2})$ are numbers of electrons (angular momenta) of 1 and 2 interacting subshells. For the Coulomb interactions in one shell, the normalization factor $N_1(N_1-1)/[(4l_1{+}2)(4l_1+1)]$ was used. Similar approach was used for normalization of weight factors in angular parts of Coulomb interaction in (1).

The experimentally measured is the differential in angle photoionization cross section $d\sigma_{nl}(\omega)/d\Omega$ that is determined by the following expression

$$\frac{d\sigma_{nl}(\omega)}{d\Omega} = \frac{\sigma_{nl}(\omega)}{4\pi}[1 + \beta_{nl}(\omega)P_2(\cos\theta) + \kappa\gamma_{nl}(\omega)P_1(\cos\theta) + \kappa\eta_{nl}(\omega)P_3(\cos\theta)]. \tag{3}$$

Here $\sigma_{nl}(\omega)$ is the $nl$-subshell absolute photoionization cross-section, $\kappa = \omega / c$, $c$ is the speed of light, $P_i(\cos\theta)$ are the Legendre polynomials, $\beta_{nl}$ is the dipole while $\gamma_{nl}$ and $\eta_{nl}$ are the non-dipole angular anisotropy parameters, $\theta$ is the angle between polarization vector and direction of photoelectron emission.

Calculations of the cross-section and angular anisotropy parameters were performed for subshells $nl = 3s; 3p$ and emission angle $\theta = 0$, using formulas from [4] modified by multiplying with presented above normalization factors. It appeared, in accord with the results of [6, 7] that the contribution of two last terms in (3) is negligible. Therefore instead of (3), considerably simpler expression was in fact measured

$$\frac{d\sigma_{nl}(\omega)}{d\Omega} = \frac{\sigma_{nl}(\omega)}{4\pi}[1 + \beta_{nl}(\omega)]. \tag{3}$$

**3. Specific of the experiments geometry.** In experiment the measurement was performed at such an angle that the really obtained quantity was, according to (3), not the photoionization cross-section $\sigma_{3s,3p}(\omega)$, but $\sigma_{3s,3p}(1+\beta_{3s,3p})$. We have calculated the cross sections and angular anisotropy parameters, dipole and non-dipole, in the frame of one electron Hartree-Fock (HF) approach and with account of RPAE multi-electron correlations. The dipole parameters as well as cross-sections proved to be almost the same in HF and RPAE, thus signaling that the role of RPAE corrections is small enough. For 3s subshell $\beta_{3s} = 2$, while for $\beta_{3p}(\omega)$ the following relation [4] was employed

$$\beta_{3p}(\omega) = \frac{D_{3p\to\varepsilon d}}{[D^2_{3p\to\varepsilon s} + 2D^2_{3p\to\varepsilon d}]}[D_{3p\to\varepsilon d} - 2D_{3p\to\varepsilon s}\cos(\delta_s - \delta_d)], \quad (4)$$

where $D_{3p\to\varepsilon d(s)}$ are the module of the dipole photoionization amplitude of the $3p \to \varepsilon d(s)$ transitions, respectively and $\delta_{d(s)}$ are the sums of amplitudes and photoelectrons scattering $d$ and $s$ scattering phases.

**4. The results obtained.** Using (3) by substituting there dipole angular anisotropy parameters, the 3s and 3p photoionization cross-sections were obtained. In [1] the $\beta$ parameters were taken from [5, 6]. Fig.1 shows the result obtained for $\sigma_{3s}(\omega)$ and $\sigma_{3p}(\omega)$, measured and calculated. On the same Fig. 1 we present results of [2] and in HF – Slater approximation [6, 7]. We see in experiment essential difference, noticeable even in logarithmic scale, between photoionization of Ni thin films and crystal samples. As to calculations, they were performed for an isolated Ni atom.

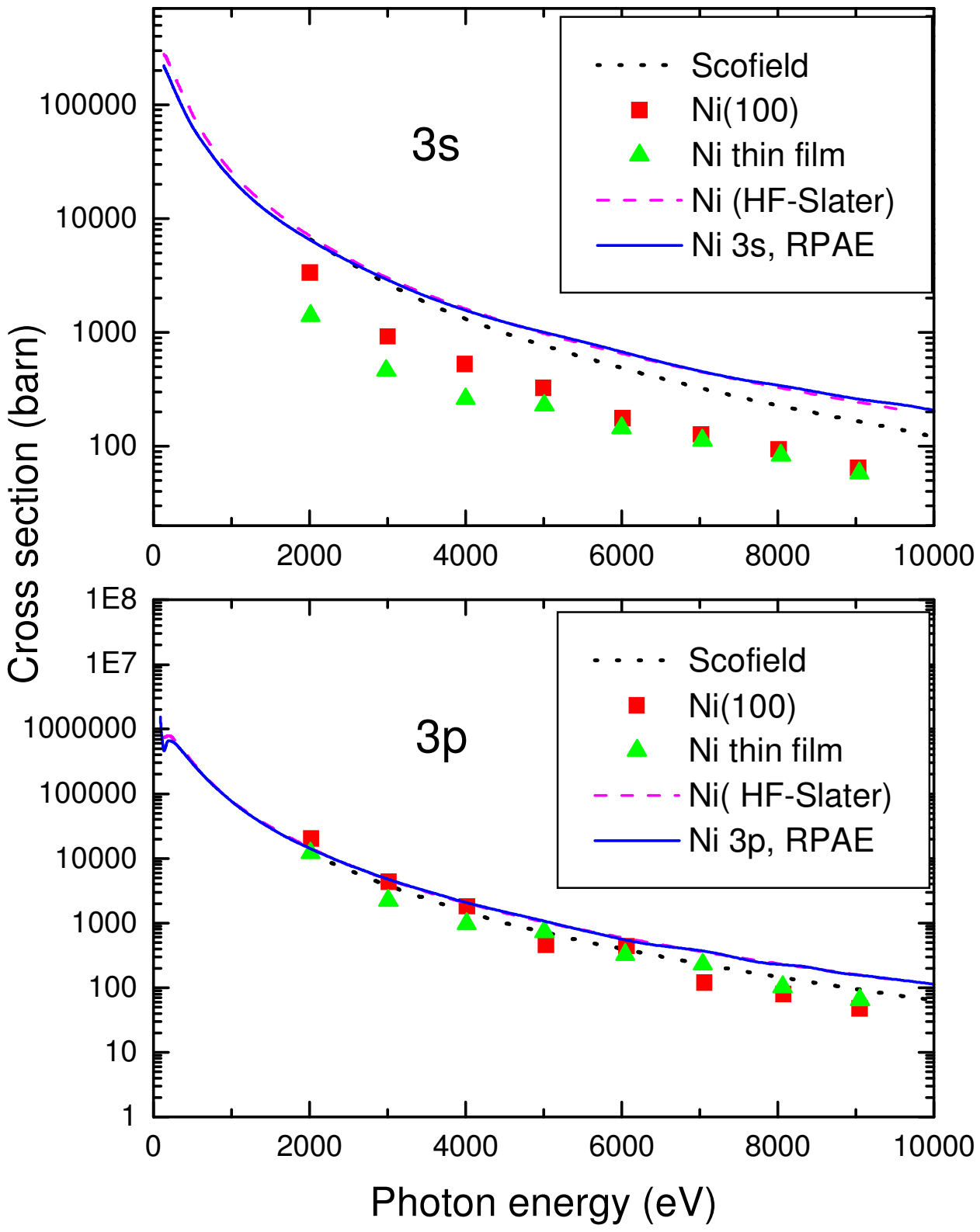

Fig. 1 The cross sections of 3s and 3p-subshells of 3s and 3p Ni subshells measured in thin films and solid body along with different calculation data.

Calculations demonstrate that in the considered photon energy range, 2-10 keV, the role of RPAE correlations is almost negligible. In the same logarithmic scale there is no difference between HF-Slater and RPAE results. The data from [2] is closer to experiment than RPAE results.

Comparison between theory and experiment demonstrates reasonable agreement for 3p subshell. For 3s subshell, the relative difference between theory and experiment is steadily and very slow increasing with the photon energy growth. This difference is big, the experimental value being smaller than the calculated one by a factor of 2.5-3.

As is seen from Fig. 1, for 3p subshell the deviation between results of calculation and measurement is much smaller than in the 3s case. In the entire considered photon energy interval, the measured cross section is smaller than the calculated and the difference slowly increases with $\omega$ growth.

Fig.2 shows the result directly observed in experiment: values of the ratio $\sigma_{3s}(1+\beta_{3s})/\sigma_{3p}(1+\beta_{3p})$ both measured [1] and calculated. On the same Figure along with our calculations we present results derived from [2] and in HF – Slater approximation [6, 7]. As is mentioned above, we put $\beta_{3s}=2$. The linear instead of logarithmic scale permits to see the differences between theory and experiment more clearly. In the entire considered photon energy region the ratio increases. Since the role of angular anisotropy parameters in the presented in Fig. 2 ratio is important but depends upon photon energy weakly, as it is seen in Fig. 3, the linear increase of $\sigma_{3s}(1+\beta_{3s})/\sigma_{3p}(1+\beta_{3p})$ in the region $\omega=1\div 10KeV$ reflects the linear increase of the ratio $\sigma_{3s}/\sigma_{3p}$ in the hydrogenlike approximation. It is seen that our results as well as that from [6, 7] are considerably higher than that of [2]. The small role of electron correlations is clearly seen.

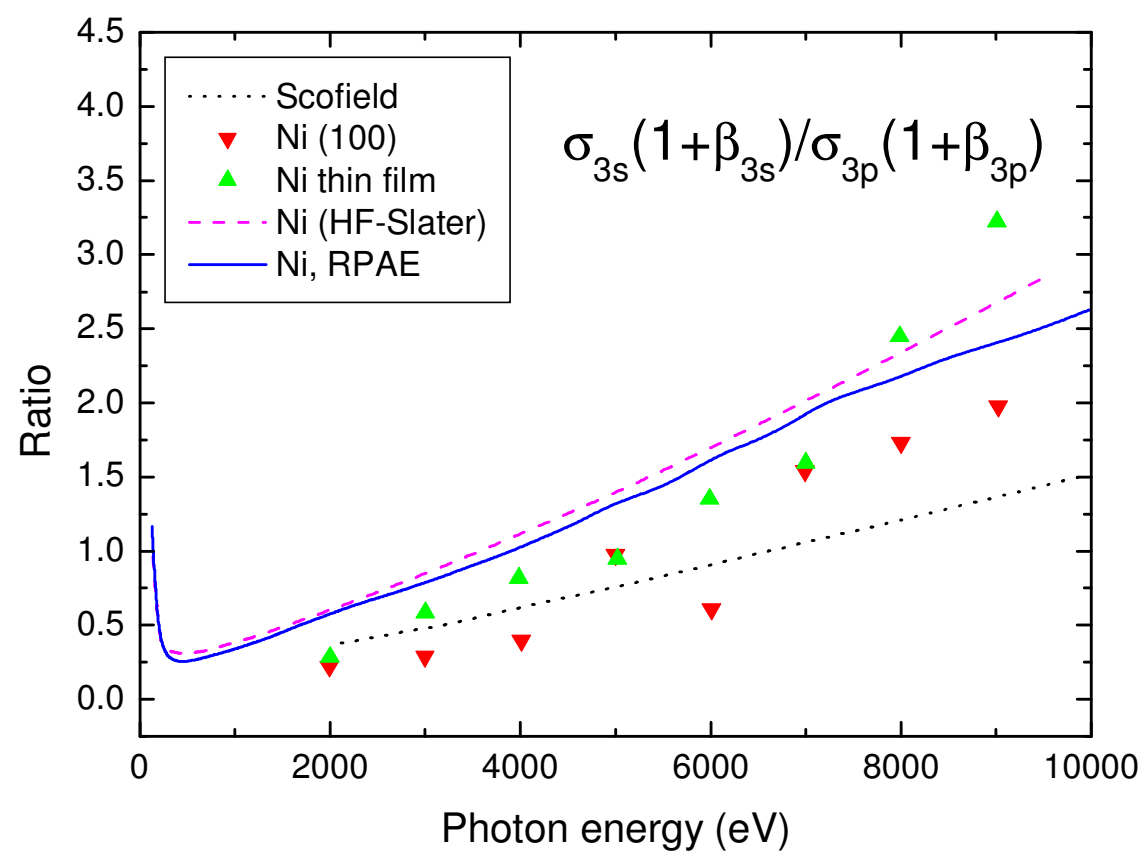

Fig. 2. The results of measurement [1] and calculation for the ratio $\sigma_{3s}(1+\beta_{3s})/\sigma_{3p}(1+\beta_{3p})$ in Ni film and crystal.

As to experimental data, they are considerably deviating from the linear low. Starting from 5 KeV the data for Ni film increases much faster than a simple, $\sim\omega$, function. The same is valid for the crystal Ni data, if imagine a smooth curve that goes with the smallest possible deviation via the experimental Ni bulk points.

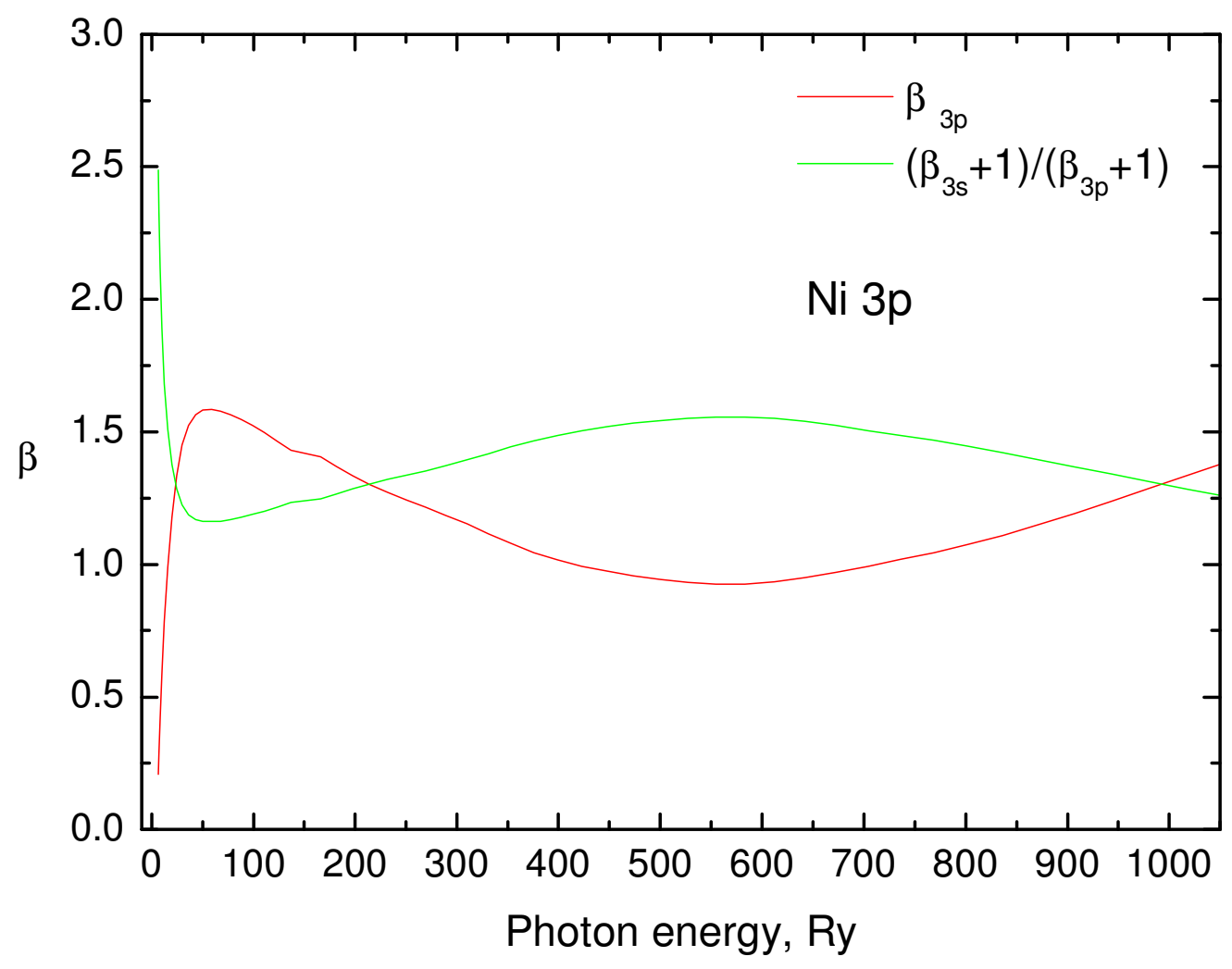

Fig. 3. Angular anisotropy parameter of Ni 3p electrons and the ratio $(1+\beta_{3s})/(1+\beta_{3p})$.

It is seen that the RPAE results are in general closer to the thin film data. The results for crystal Ni up to 6 KeV are closer to the data of [2] than to RPAE and only at higher $\omega$ lie closer to RPAE. Entirely, the linear scale emphasizes deviation between calculation and experimental data even more clear than the logarithmic scale of Fig. 1.

As is seen from Fig. 3, the increase of $\beta_{3p}$ is fast below $\approx 800eV$, but in the whole considered photon energy region $\omega=2-15keV$ it varies slowly from 1.5 to about 1 and back, having a broad shallow minimum. As to the cross sections they decrease in the same photon energy region monotonically by several orders of magnitude.

**5. Concluding remarks.** Neither HF nor RPAE are able to describe the experimental data for the absolute values of the cross-sections or their ratio, as is evident from Fig. 1 and 2. Since the dynamic correlations rapidly die out with photon energy increase, it is hard to believe that these correlations can be responsible for the observed difference.

As it was demonstrated in [3], there exist some RPAE corrections that could be of importance at high $\omega$. However these are not affecting 3s cross section and in principle could increase the 3p cross section, which is already too big.

Of importance could be the effect of the so-called spectroscopic factor $F_{nl} < 1$ of a $nl$ level, even of an s-level [9]. This factor represents an admixture due to electron correlations of other level or of the continuous spectrum to the $nl$ level under consideration. It modifies the cross- section simply by multiplying its RPAE value by $F_{nl}$. Thus, in order to achieve agreement with experimental data one needs $F_{3s} \approx 0.3$ that is unlikely. At least direct calculation for Ni neighbor, Mn atom, gives for $F_{3s}$ a considerably bigger value close to 0.7 [9], thus ruling out that the big difference between theory and experiment could be explained by inclusion of the spectroscopic factor. Note, that the measured in [1] structure of 3s level also excludes the smallness of $F_{3s}$ and its important role.

At first glance, a little bit suspicious is the limitation in the angular distribution (3) by dipole and quadrupole corrections only. Indeed, the corresponding multipole expansion parameter can be estimated as 0.2. It is not too small but makes the large contribution of the term of second power in $\kappa$, neglected in (3), unlikely. This conclusion is confirmed by calculations in [7] that included the $\kappa^2$ terms, which proved to be three orders of magnitude smaller.

It surprises also the strong influence of the shape of the sample, namely, whether it is a crystal or a film. For levels with binding energy of 110 eV (for 3s) and 67 eV (for 3p) a strong effect of location of neighboring atoms looks almost improbable.

Entirely, the investigation of atomic photoionization in the tens keV region requires and deserves further efforts and clarifications, both theoretical and experimental.